\documentstyle[12pt,epsfig]{article}
\input epsf
\newcommand{\be}[1]{
\begin{eqnarray}\label{#1}}
\newcommand{\ee}{\end{eqnarray}}

\newcommand{\re}[1]{(\ref{#1})}

  \newcommand{\gsim}
  {\mbox{\hspace*{0.1em}\raisebox{.4ex}{$\scriptstyle <$}
  \hspace{-0.88em}\raisebox{-.6ex}{$\scriptstyle\sim$}\hspace*{0.05em}}}

\def\kslash{\rlap/{\mkern-1mu k}}

\begin{document}

\renewcommand{\thefootnote}{\fnsymbol{footnote}}
\begin{flushright}
\begin{tabular}{l}
RUB-TP2-01/01\\
\end{tabular}
\end{flushright}
\begin{center}
{\bf\Large Soft pion theorem for hard near threshold pion
production}

\vspace{0.2cm}

\begin{center}
P.V.~Pobylitsa$^{a,b}$, M.V.~Polyakov$^{a,b}$, M.~Strikman$^{a,c}$\\[0.1cm]
{$^a$Petersburg Nuclear Physics Institute, 188350, Gatchina,
Russia\\ $^b$ Institut f\"ur Theoretische Physik II,
Ruhr-Universit\"at Bochum,\\ D-44780 Bochum, Germany\\
$^c$ Pennsylvania State University, University Park, PA 16802, USA }

\end{center}
\end{center}
\vspace{0.1cm}

\begin{abstract}
We prove a new soft pion theorem for the near threshold pion production
by a hard electromagnetic probe. This theorem relates
various near threshold pion production amplitudes to the nucleon
distribution amplitudes.
The new soft pion theorem
 is in a good agreement with the SLAC data for the
 structure function  $F_{2}^p(W,Q^2)$
for $W^2\le 1.4$~GeV$^2$ and $7\le Q^2\le 30.7$ GeV$^2$.
\end{abstract}

\section*{\normalsize \bf Introduction}
The amplitudes of the pion production near the threshold by the electromagnetic  probe
\be{DIS}
\gamma^*+N \to \pi+N'
\ee
at a not too large virtuality ($Q^2$) of the photon
$Q^2 \ll \Lambda^3/m_\pi$ ($\Lambda\sim 1$~GeV is a typical hadronic scale)
can be expressed in terms of various nucleon form factors with the
help of the soft pion theorem (SPT) \cite{old,VZ,Koch}.
For virtualities $Q^2 \sim \Lambda^3/m_\pi$
and larger this SPT
does not work~\cite{old,VZ,Koch}.

In the present paper we derive a new ``hard-soft" pion theorem
(hSPT)
for the reaction \re{DIS} in the near threshold region
and for
large virtuality of the photon $Q^2\gg \Lambda^2$.
Our main tool is
the QCD factorization theorem for exclusive processes
\cite{BL,ER} (for a recent review and comprehensive list of references see e.g.
\cite{Brodsky:2000dr}).
It allows us to
express the pion production amplitude at large virtuality in terms
of the distribution amplitudes (DAs)
of the nucleon and of the low-mass $\pi N$ system.
These non-perturbative objects correspond to the lowest
three quark (3q)
Fock component of the nucleon and $\pi N$ systems.
We derive a  hSPT
to relate  the corresponding
 distribution amplitude of the $\pi N$
system (we shall call it as $\pi N$~DA)
to the nucleon distribution amplitude.
This hSPT is valid
for the limit
 when the mass of the $\pi N$
system (denoted as $W$) is close to the threshold $W_{\rm
th}=M_N+m_\pi$, {\em i.e.}
$W-W_{\rm th}\gsim m_\pi$. The derivation of the similar
theorem for DA of the two pion system near threshold can be found in \cite{Ter,MVP98}.

The physical picture of the near threshold production of pion by
a hard
electromagnetic probe is as follows. At large $Q^2$ the emission of
the soft pion from the initial state contributes only to
large invariant masses $W$. The emission from the hard interaction
part is a higher twist in $Q^2$.  Hence the emission occurs solely
in the final state when a small $3q$ system produced in the hard
interaction expands to a large enough configuration. At this point
we are dealing with a soft pion emission and can apply
corresponding near threshold chiral theory relations.

\section*{\normalsize \bf Threshold theorem for hard electromagnetic probe}
We consider the  hard reaction \re{DIS}
in the near threshold region, $i.e.$ when $W$ is close to the
threshold of the reaction $W_{\rm th}=M_N+m_\pi$ and
at large virtuality of the photon $Q\gg \Lambda$.
Hence,
$x_{Bj}\to \left[1+m_\pi(2
M_N+m_\pi)/Q^2\right]^{-1}$ in the discussed limit.
In the near threshold region the
cross section of the reaction \re{DIS} can be expressed in terms
of the $N\to \pi N$ transition form factors at large momentum transfer
$Q^2$:

\be{NpiN}
M_\mu^{ff'a}=
\langle \pi^a N_{f'}| J_\mu^{\rm e.m.}(0)|N_f\rangle
\ee
Here $f,f'=p,n$ are flavours
of the initial and final nucleons, $a$ is the
flavour index of the emitted pion.

In this section we shall be interested in the matrix element \re{NpiN}
at the threshold of the reaction \re{DIS}, i.e.
 at $W=W_{\rm th}$.  The matrix element \re{NpiN} at the threshold
 for the transverse photon has the form\footnote{Longitudinally polarized
 photon gives a power suppressed contribution for $Q^2\gg \Lambda^2$.}:

\be{NpiNth}
M_{\mu_\perp}^{ff'a}= A(\gamma^* N_f \to N_{f'}+\pi^a)\
\bar u(p')\gamma_{\mu_\perp}\gamma_5 u(p)\, .
\ee

It follows from the
QCD factorization theorem \cite{BL,ER}
that
the transition matrix element $A(\gamma^* N\to \pi N')$
at large $Q^2$
can be written as
(up to the
power suppressed terms)
\be{fac}
A(\gamma^* N\to \pi N') = \int dx dy \Phi_{\pi N'}^* (x) T(x,y,Q^2)
\Phi_N(y)\, ,
\ee
where $T(x,y,Q^2)$ is the hard part of the process  computable in
perturbative QCD. The functions
$\Phi_{N}(y)$, $\Phi_{\pi N'}(x)$ are
distribution amplitudes
(light cone wave functions)
 of the nucleon and of the low-mass $\pi N'$
system. The DA of the nucleon $\Phi_{ N}(y)$ also enters the QCD
description of the nucleon form factor and it is a subject
of intensive studies. The
distribution
amplitude of the $\pi N$ system is
a straightforward
generalization of
the baryon DAs.
Many general
properties of these objects are similar to
those
of the two-pion distribution amplitude which were
extensively studied
in Ref.~\cite{MVP98}. They will be discussed elsewhere.

We
focus here on the soft pion theorem for $\pi N$~DAs.
Using the
general soft pion theorem
(see e.g. \cite{kniga}) we can write:
\begin{equation}
\label{spgeneral}
\langle \pi^a(k) N_f(p,S)| O | 0 \rangle =
\frac{i}{f_\pi} \langle N_f(p,S)| \left[ Q_5^a, O\right]|0\rangle
\end{equation}
\[
+
\frac{i g_A}{4 f_\pi (p\cdot k)} \sum_{S',f'} \bar u(p,S) \kslash
\gamma_5 \tau^a_{ff'}u(p,S')
\langle N_{f'}(p,S')| O | 0\rangle \, .
\]
Here $f_\pi=93$~MeV is the pion decay constant,
$g_A\approx 1.25$ is the axial charge of the nucleon. The operator $O$ is
the trilocal quark operator of twist-3 in which
quark fields are separated  by a light cone distance, in the complete
analogy with the definition of the baryon DA, see e.g. \cite{CZ}.
The
operator $Q_5^a$ is the operator of axial charge, so that the
calculation of the
commutator
$\left[ Q_5^a, O\right]$ reduces to the chiral rotation of
operator $O$.
 The second term in the rhs of Eq.~(\ref{spgeneral}) corresponds to
 the chiral singularity due to the nucleon pole in the graph
shown in Fig.~1.
The contribution of this diagram is strongly suppressed  for $W-W_{\rm th}\ll m_{\pi}$
but for $W-W_{\rm th}\sim  m_{\pi}$ it becomes significant,
see Eqs.(\ref{spf2p},\ref{spf2n}).
\begin{figure}[h]
\epsfxsize=4cm \centerline{\epsffile{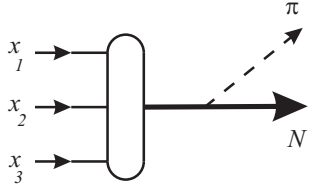}} \caption{Nucleon
pole contribution to the soft pion theorem for generalized $\pi N$
distribution amplitudes. } \label{iaa}
\end{figure}

Let us start from the calculation of the first (commutator) term
in Eq.~(\ref{spgeneral}).
Since the chiral rotation of the trilocal quark operator $O$ does not change
its twist Eq.~(\ref{spgeneral}) allows us to express generalized $\pi
N$ DAs
at the threshold
 in terms of
nucleon DAs.

We write the nucleon DA in terms of functions $\phi_S(x)$
and $\phi_A(x)$ which are symmetric and antisymmetric
respectively under $x_1 \leftrightarrow x_3$ (1 and 3 are quarks with parallel
helicities) \cite{BL,CarPoo}. For the proton we have
\be{wfp}
\nonumber
|p \uparrow\rangle&=&\frac{\phi_S(x)}{\sqrt 6}\
|2 u_\uparrow d_\downarrow u_\uparrow -u_\uparrow u_\downarrow d_\uparrow
-d_\uparrow u_\downarrow u_\uparrow\rangle\\
&+& \frac{\phi_A(x)}{\sqrt 2} |u_\uparrow u_\downarrow d_\uparrow
-d_\uparrow u_\downarrow u_\uparrow
\rangle\, .
\ee
The distribution amplitude for neutron can be obtained from the above
expression by the replacement $u \leftrightarrow d$.

Applying the general soft pion theorem (\ref{spgeneral})
we express
the distribution amplitudes of various $\pi N$
systems at the threshold in terms of the nucleon DAs $\phi_S(x)$
and $\phi_A(x)$:
\be{wfpp0}
\nonumber
|p \uparrow \pi^0 \rangle&=&\frac{\phi_S(x)}{2 \sqrt 6 f_\pi}\
|6 u_\uparrow d_\downarrow u_\uparrow +u_\uparrow u_\downarrow d_\uparrow
+d_\uparrow u_\downarrow u_\uparrow\rangle\\
&-& \frac{\phi_A(x)}{2  \sqrt 2 f_\pi} |u_\uparrow u_\downarrow d_\uparrow
-d_\uparrow u_\downarrow u_\uparrow
\rangle\, ,
\ee
\be{wfnpp}
\nonumber
|n \uparrow \pi^+\rangle&=&\frac{\phi_S(x)}{\sqrt{12} f_\pi}\
|2 u_\uparrow d_\downarrow u_\uparrow -3\ u_\uparrow u_\downarrow d_\uparrow
-3\ d_\uparrow u_\downarrow u_\uparrow\rangle\\
&-& \frac{\phi_A(x)}{2 f_\pi } |u_\uparrow u_\downarrow d_\uparrow
-d_\uparrow u_\downarrow u_\uparrow
\rangle\, .
\ee
The DAs of the neutral $\pi N$ systems can be obtained by the trivial replacement
$u \leftrightarrow d$.

Now we can compute
the threshold amplitudes
$A(\gamma^* N\to \pi N')$
at large $Q^2$ combining the factorization theorem (\ref{fac})
with the expressions for $\pi
N$~DAs (\ref{wfpp0},\ref{wfnpp}).
The technique of calculations of the hard part $T(x,y,Q^2)$ is
standard and can be found, e.g. in refs.~\cite{BL,CZ}.
We give here only the final results using notations of refs.~\cite{BL,CarPoo,CGS}.
For the transitions $p\to \pi N$ we obtain:
\be{resup0p}
Q^4 A(\gamma^* p\to \pi^0 p)|_{{\rm th}}&=& \frac{(16\pi \alpha_s)^2}{9 f_\pi}\
\int [dx\ dy]\ \biggl\{
\frac 1 9 \biggl(23 T_1-8 T_2 \biggr)\ \phi_S(x) \phi_S(y)\\
&+&
\nonumber
\frac{1}{\sqrt 3} T_1
\biggl[\phi_S(x) \phi_A(y) + \phi_S(y) \phi_A(x)\biggr] \\
\nonumber
&-&\frac 13 \biggl(T_1+2 T_2 \biggr)\phi_A(y) \phi_A(x)
\biggr\}\, ,
\ee

\be{resuppn}
Q^4 A(\gamma^* p\to \pi^+ n)|_{{\rm th}}&=& \frac{(16\pi \alpha_s)^2}{9 \sqrt 2 f_\pi}\
\int [dx\ dy]\ \biggl\{
\frac 2 9 \biggl(11 T_1+4 T_2 \biggr)\ \phi_S(x) \phi_S(y)\\
&-&
\nonumber
\frac{2}{\sqrt 3} T_1
\biggl[\phi_S(x) \phi_A(y) + \phi_S(y) \phi_A(x)\biggr] \\
\nonumber
&-&\frac 23 \biggl(T_1+2 T_2 \biggr)\phi_A(y) \phi_A(x)
\biggr\}\, .
\ee
Here $[dx]=dx_1 dx_2 dx_3\ \delta(1-x_1-x_2-x_3)$ and coefficient
functions $T_1$ and $T_2$ have the form \cite{BL}:
\be{t1t2}
T_1&=&\frac{1}{x_3(1-x_1)^2y_3(1-y_1)^2}+\frac{1}{x_2(1-x_1)^2y_2(1-y_1)^2}
\\
\nonumber
&-& \frac{1}{x_2 x_3(1-x_3)y_2 y_3(1-y_1) }\, ,\\
T_2&=&\frac{1}{x_1 x_3 (1-x_1)y_1 y_3(1-y_3)}\, .
\ee
Expressions for various nucleon form factors in the same notations
can be found e.g. in refs.~\cite{BL,CarPoo,CGS}.

For the transitions $n\to \pi N$ we obtain:
\be{resup0n}
Q^4 A(\gamma^* n\to \pi^0 n)|_{{\rm th}}&=& \frac{(16\pi \alpha_s)^2}{9 f_\pi}\
\int [dx\ dy]\ \biggl\{
\frac{13} 9 \biggl(T_1-T_2 \biggr)\ \phi_S(x) \phi_S(y)\\
&+&
\nonumber
\frac{1}{\sqrt 3} T_1
\biggl[\phi_S(x) \phi_A(y) + \phi_S(y) \phi_A(x)\biggr] \\
\nonumber
&+&\frac 13 \biggl(T_1- T_2 \biggr)\phi_A(y) \phi_A(x)
\biggr\}\, ,
\ee
\be{resupmp}
Q^4 A(\gamma^* n\to \pi^- p)|_{{\rm th}}&=& \frac{(16\pi \alpha_s)^2}{9 \sqrt 2
f_\pi}\ \int [dx\ dy]\ \biggl\{ -\frac 2 9 \biggl(T_1-T_2 \biggr)\
\phi_S(x) \phi_S(y)\\
\nonumber
&+& \frac{2}{\sqrt 3} T_1 \biggl[\phi_S(x)
\phi_A(y) + \phi_S(y) \phi_A(x)\biggr] \\
\nonumber
&-&\frac 23 \biggl(T_1- T_2
\biggr)\phi_A(y) \phi_A(x) \biggr\}\, .   \ee

Using
Eq.~(\ref{resup0p}) one can express the near threshold
structure functions $F_2^{p,n}(W,Q^2)$
and various  differential
cross sections for
 particular $\pi N$ channels
in terms of the nucleon DAs.
The {\em same } DAs appear in the QCD factorization theorem for
the nucleon form factors at large $Q^2$.
We find that in the case of the symmetric form of the nucleon DA
\be{eta}
\phi_S(x)\ {\rm\ is\ arbitrary}\, , \quad \phi_A(x)=0\, ,
\ee
one can describe the near threshold pion production directly in terms of the nucleon form
factors without specifying the nucleon DA, see below.
Certainly the nucleon DA  can have a nonzero asymmetric component $\phi_A(x)$.
Actually our general Eq.~(\ref{resup0p}) can be applied to any specific
model of the nucleon DA (see e.g. \cite{models}) to compare the
model predictions with the experiment.
In the case of symmetric nucleon DA
the amplitude of the process
(\ref{DIS}) can be expressed in terms of nucleon
magnetic form factors ($G_{MN}(Q^2)$) as follows:
\be{etaresultsp}
A(\gamma^*p\to \pi^0 p)|_{{\rm th}}&=&
-\frac{1}{f_\pi}\left(\frac 5 6\ G_{Mp}-\frac 4 3\ G_{Mn} \right)\, ,\\
\label{etaresultsp1}
A(\gamma^*p\to \pi^+ n)|_{{\rm th}}&=&
\frac{1}{\sqrt 2 f_\pi}
\left(\frac 5 3\ G_{Mp}+\frac 4 3\ G_{Mn} \right)
\, ,
\ee
\be{etaresultsn}
A(\gamma^*n\to \pi^0 n)|_{{\rm th}}&=& -\frac{13\ G_{Mn}}{6
f_\pi}\, ,\\
A(\gamma^*n\to \pi^- p)|_{{\rm th}}&=&
\frac{G_{Mn}}{3 \sqrt 2 f_\pi}\, .  \ee
Since these
results
are based on
the symmetric form \re{eta} for the nucleon distribution
amplitude,  deviations from these equations
 would allow to probe
directly the asymmetric part of the nucleon distribution
amplitude
and check the validity of the leading twist description of the nucleon
form factors.

At asymptotically large $Q^2$ the DAs of the
nucleon are evolved to their asymptotic values:
\be{asyda}
\phi_S^{\rm as}(x)=N\ x_1 x_2 x_3\, , \quad \phi_A^{\rm as}(x)=0\, .
\ee
As the form (up to overall normalization) of the asymptotic
nucleon DAs is fixed uniquely
we can make predictions for
the ratios of various amplitudes at the threshold at $Q^2\to \infty$
$
\lim_{Q^2\to \infty}\
\frac{A(\gamma^*p\to \pi^0 p)|_{{\rm th}}}{
A(\gamma^*p\to \pi^+ n)|_{{\rm th}}}
= -\sqrt 2 + O\biggl(\frac{m_\pi}{M_N}\biggr)$,
$\lim_{Q^2\to \infty}\
\frac{A(\gamma^*n\to \pi^0 n)|_{{\rm th}}}{
A(\gamma^*n\to \pi^- p)|_{{\rm th}}}
= -\frac{13}{\sqrt 2} + O\biggl(\frac{m_\pi}{M_N}\biggr)$
which are qualitatively different from predictions for the case $Q^2\to 0$ \cite{old}
$
\lim_{Q^2\to 0}\
\frac{A(\gamma^*N\to \pi^0 N)|_{{\rm th}}}{
A(\gamma^*p\to \pi^+ n)|_{{\rm th}}}
=  O\biggl(\frac{m_\pi}{M_N}\biggr)
$ For the ratio of neutron to proton structure functions we obtain from
new soft pion theorems the following asymptotic result:
$
\lim_{Q^2\to \infty, W\to W_{{\rm th}}}
\frac{F_2^{n}(W,Q^2)}{F_2^{p}(W,Q^2)}=\frac{57}{32}
+ O\biggl(\frac{m_\pi}{M_N}\biggr)\, ,
$
which is much larger than the perturbative QCD scaling limit
expectation of 3/7 \cite{FJ}. Although it is worth noting that
the presently experimentally accessible values
of $Q^2$ are not very large.  Therefore  one
may expect considerable
deviations from the above asymptotic results.

\section*{\normalsize \bf Structure functions: comparison with data}

The data on the structure function $F_2^p(W,Q^2)$ in the near threshold region were
obtained in 1994
in the SLAC experiment E136 \cite{Bosted}
for a wide
range of $Q^2=7\div 30.7$~GeV$^2$.
Here
we make
the first
comparison of the data with the hSPT predictions.
(Previous analyses addressed the scaling features of
these data and did not attempt to calculate
$F_2^p(W\le 1.2 GeV,Q^2)$.)

To compute
$F_2^p(W,Q^2)$ for $W-W_{\rm th}\gsim
m_\pi$ we combine the strictly threshold amplitude (\ref{resup0p})
with the contribution of the last term from Eq.~(\ref{spgeneral})
and obtain

\be{comm}
F_2^p(W,Q^2)= \frac{Q^2\beta(W)}{(4\pi)^2}\Biggl[\sum_{X=p\pi^0,\ n \pi^+} \left|
A(\gamma^* p\to X)|_{\rm th}\right|^2\\
+
\nonumber
\frac{3 g_A^2G_{M p}^2(Q^2) \beta^2(W) W^4
}{4(W^2-M_N^2+m_\pi^2)^2}+
O\biggl(\frac{m_\pi}{\Lambda}\biggr) \Biggr]\, ,
\ee
where
\be{beta}
\beta(W)=\sqrt{1-\frac{(M_N+m_\pi)^2}{W^2}}\
\sqrt{1-\frac{(M_N-m_\pi)^2}{W^2}}\, .
\ee
The first term in the rhs of eq.~(\ref{comm}) corresponds to
strictly threshold amplitude (\ref{resup0p}).
The second term takes into account the
emission of the pion from the outgoing nucleon (Fig.~1) with the
amplitude given by the second term in the rhs of
eq.~(\ref{spgeneral}).
The latter contribution vanishes exactly at the threshold
but gives a parametrically unsuppressed contribution for $W-W_{\rm th}\sim
m_\pi$.
Note that the second term in Eq.~(\ref{comm}) corresponds to $\pi N$
system in P-wave, therefore it can be separated from the first
(S-wave)
term by considering the angular distributions in $\pi N$ system.

The data of E136 experiment
Ref.~\cite{Bosted} are consistent with the factorized form
\be{fit}
F_{2}^p(W,Q^2)=F(Q^2)\ G(W),
\ee
with $F(Q^2)\propto 1/Q^6$ for $Q^2\ge 8$~GeV$^2$ which
is exactly the scaling behaviour following from Eq.~(\ref{comm}).
We also
found  that  Eq.~(\ref{comm})
provides a good description of the W-dependence of the E136 data
for $W\le 1.2 $~GeV
though the predicted  $W$-dependence
is somewhat different from the $G(W)\propto W^2 -W_{\rm th}^2$
fit of \cite{Bosted}. However, the resolution of E136 is not sufficient
to distinguish between the two forms.
The hSPT (\ref{comm}) also predicts the scaling
behaviour $\sim 1/Q^6$ which is confirmed by the E136
data \cite{Bosted}.

Thus we reproduce several features of the data without
using any specific nucleon wave functions.
To make a
first
 quantitative
comparison of the soft pion theorem
prediction for the absolute value of $F_2^p$
we use the symmetric form for the
nucleon distribution amplitude \re{eta}.
Inserting the expressions (\ref{etaresultsp}) and
(\ref{etaresultsp1}) for the threshold amplitudes into Eq.~(\ref{comm})
and using the analogous expression for the neutron structure function, we
obtain the following hSPT for the structure
functions in the case of the symmetric form (\ref{eta}) for
the nucleon
DA

\be{spf2p}
F_2^p(W,Q^2)&=&\frac{Q^2 \beta(W)}{(4\pi f_\pi)^2}\Biggl[
\frac{25}{12} \ G_{Mp}^2(Q^2) +\frac 83 \ G_{Mn}^2(Q^2)
\\
&+&
\nonumber
\frac{3 g_A^2G_{M p}^2(Q^2) \beta^2(W) W^4 }{4(W^2-M_N^2+m_\pi^2)^2}
+O\biggl(\frac{m_\pi}{\Lambda}\biggr) \Biggr]\, ,
\ee
\be{spf2n}
F_2^n(W,Q^2)&=&\frac{Q^2 \beta(W)}{(4\pi f_\pi)^2} \Biggl[\frac{19}{4}\ G_{Mn}^2(Q^2)
\\
&+&
\nonumber
\frac{3 g_A^2G_{M n}^2(Q^2) \beta^2(W) W^4 }{4(W^2-M_N^2+m_\pi^2)^2}\
+
O\biggl(\frac{m_\pi}{\Lambda}\biggr) \Biggr]\, .
\ee

Now we can compare our results based on the
hSPT with the E136
data~\cite{Bosted}. It was found in this experiment
that in the interval $W^2\le 1.4$~GeV$^2$ the quantity $\int_{\rm
th}^{1.4} dW^2 Q^6 F_2^p(W,Q^2)$
is practically constant
for $Q^2\ge 8$~GeV$^2$.
The experimental value of the integral is in a good
agreement  with
our result (\ref{spf2p}):
\be{sp}
\int_{\rm th}^{1.4}\hspace{-1em}
dW^2 Q^6 F_2^p(W,Q^2) =\left\{
\begin{array}{l}
0.10 \pm 0.02\ {\rm GeV}^8\ \ ({\rm E136}) \\
0.11\pm 0.02\ {\rm GeV}^8\ \ ({\rm hSPT})
\end{array}
\right.
\ee
For the theoretical analysis we used as input the following values
for the nucleon form factors
$Q^4 G_{Mp}(Q^2)= 1.0\pm 0.1$~GeV$^4$ obtained at $Q^2\ge
10$~GeV$^2$ \cite{expff} and $Q^4 G_{Mn}(Q^2)=-(0.5\pm 0.1)$~GeV$^4$
extracted from Ref.~\cite{expffn} at $Q^2\approx 10$~GeV$^2$
\cite{komentarij}.
For the summary of the current information on
various baryon form factors at large momentum transfer see e.g.
\cite{Stoler:1993yk}.
Note that the contribution of the P-wave
term in Eq.~(\ref{spf2p}) is relatively small (about 20\%), so
that the main part of the hSPT value (\ref{sp}) is due to the
strictly threshold contribution (S-wave) in
Eq.~(\ref{comm}).
With the same set of parameters we
predict
$
\int_{\rm th}^{1.4} dW^2 Q^6 F_2^n(W,Q^2)= 0.05\pm 0.02 ~{\rm
GeV}^8
$.
Note that
for the nucleon DAs which fit $G_{Mp}(Q^2),G_{Mn}(Q^2)$ at $Q^2\ge 10$~GeV$^2$
(such DA significantly differ from the
asymptotic one, see e.g. \cite{CarPoo,CGS})
 the $F_2^n/F_2^p$ ratio
near threshold
is much smaller than the asymptotic
value of $52/37$
 which follows from Eq.(\ref{resup0p})
with the asymptotic distribution amplitude $\phi(x)\propto x_1x_2x_3$.
Thus this ratio is
extremely sensitive to the
deviations of the nucleon DA from the asymptotic form.
Therefore measurements of the neutron structure function in the near
threshold region would considerably constrain the form of the nucleon
distribution amplitude.

\section*{\normalsize \bf Conclusions}

In this paper we derived a new soft pion theorem for the
threshold pion production by a hard electromagnetic probe, i.e.
with the probe of virtuality $Q^2 \gg \Lambda^2$ ($\Lambda\sim 1$~GeV is a
typical hadronic
scale). This new
hSPT
allows us to express the pion
production amplitudes in terms of the
distribution amplitudes of the
nucleon. The latter enter the description of various nucleon form
factors at large momentum transfer.
These new relations
give a possibility to constrain
further
the
nucleon distribution amplitude using data on
threshold inelastic electron scattering from the nucleon at high momentum
transfer.

Here we restricted ourselves only to the case of the leading chiral
contributions to the corresponding threshold amplitudes. The chiral
corrections can be rather easily computed using methods of
the
chiral
perturbation theory.

Using
a generic
symmetric model for the nucleon DAs
we demonstrated that various observables for near threshold pion
production at high momentum transfer are sensitive to the
parameters of nucleon DAs. This shows that the near threshold pion production by a
hard electromagnetic probe is a new valuable source of information about
nucleon distribution amplitudes. Studies with
a broader range of
models of
nucleon DAs will be presented elsewhere.

Our analysis was restricted to the leading twist QCD
contributions. The application of the methods developed here to the
models for soft contributions to the baryon form factors (see a review in
\cite{Radyushkin,Kroll})
would allow one to derive predictions of these models
for hard near threshold pion production.
This might be an exciting possibility to use hSPT to discriminate
between soft and hard mechanisms for high momentum transfer reactions.

We also note that the generalized $\pi N$
distribution amplitudes above the threshold
are complex functions. Not very far from the threshold the phases
of $\pi N$~DAs are fixed by known elastic $\pi N$ scattering phase
shifts which allows us to use the Omn\`es solution to
the
 corresponding
dispersion relations in order to fix the dependence of $\pi N$~DAs
on the mass of the corresponding $\pi N$ system
\cite{MVP98,MVPB98}.
Since
this dependence is fixed to
a large
extent by the $\pi N$ scattering phase shifts, the generalized
$\pi N$~DAs also contain information about DAs of various $\pi N$
resonances.
In a complete analogy with the
two-pion distribution amplitudes the generalized $\pi N$~DAs give
a possibility to describe resonance and non-resonance contributions to
the reaction~\re{DIS} at high photon virtuality in a
unified way.

The formalism developed here can be also used for other hard
reactions with the pion emission near the threshold. For instance one
can analyze the Compton scattering $\gamma +N \to \gamma+ (\pi N)$
at large momentum transfer.

The study of the discussed processes should be feasible at the top of the
current JLab energies and should be one of the
high priorities of JLab at 12 GeV.

{ We are grateful to P.~Bosted, D.~I.~Diakonov, A.~V.~Efremov,
L.~Frankfurt, K.~Goeke, V.~Yu.~Petrov, M.~Praszalowicz, A.~V.~Radyushkin,
N.~G.~Stefanis, P.~Stoler and M.~Vanderhaeghen for interesting
discussions. This research has been supported by DFG, BMFB,
A.~v.~Humboldt foundation and DOE. }

\end{document}